\documentstyle[12pt]{article}
\newcommand{\be}{\begin{equation}}
\newcommand{\ee}{\end{equation}}

\newcommand{\bea}{\begin{eqnarray}}
\newcommand{\eea}{\end{eqnarray}}

\begin{document}
\title{Constraints on  non-minimally  coupled curved space electrodynamics
from astrophysical observations}
\author{
A.R.Prasanna and  Subhendra Mohanty\\
 Physical Research Laboratory,\\
Navrangpura, Ahmedabad - 380 009, India}

%\address{ Physical Research Laboratory,
%Navrangpura, Ahmedabad - 380 009, India}

%\date{\today}
%\date{}

\maketitle

\begin{abstract}
We study interactions of electro-magnetic fields with the
curvature tensor of the form $\lambda R_{\mu \nu \alpha
\beta}F^{\mu \nu}F^{\alpha \beta}$. Such coupling terms though are
invariant under general coordinate transformation and CPT, however
violate the Einstein equivalence principle. These couplings do not
cause any energy dependent dispersion of photons but they exhibit
birefringence. We put constraints on the coupling constant
$\lambda$ using results from solar system radar ranging
experiments and millisecond-pulsar observations. We find that the
most stringent constraint comes from pulsar observations and is
given  by $ \lambda < 10^{11} cm^2 $ obtained from the timing of
binary pulsar PSR B1534+12..
\end{abstract}

\maketitle

\section{Introduction}

Einstein's general relativity is based on the twin principles of
general coordinate invariance and the Equivalence principle. A
generalization of the standard minimal coupling of
electro-magnetic fields in curved space-time introduced in
\cite{prasanna}
\begin{equation}
 {\cal L} = - \frac{1}{4}~F^{\mu \nu} ~F_{\mu \nu} +
\lambda~R^{\mu \nu \rho \sigma} ~F_{\mu \nu} F_{\rho \sigma}
\label{L}
\end{equation}

is a theory which violates the Einstein's equivalence principle
(which states that in a local inertial frame the theory should
reduce to the form given by special relativity) as the Riemann
curvature tensor does not vanish in a local inertial frame. This
theory however is invariant under general coordinate
transformation as well as CPT.

In this paper we study some of the experimental consequences of
the curvature couplings of electro-magnetic fields and put
constraints on the coupling constant $\lambda$ from  radar ranging past the sun
\cite{radar} and from the observations of
millisecond pulsars
\cite{pulsar}. The effect of the curvature term in the lagrangian
is to make the photon orbits polarization dependent. Therefore the
two independent polarizations of the electromagnic waves
experience different amounts of bending and time delay in a
gravitational field. There is however no energy dependent
dispersion of the photon trajectories.

Constraints on the
dispersion relations of the generic form $p^2-E^2 =(E/M)^n$
can be obtained from the non observation of time lag between GRB
signals of different energies which have propagated over cosmological
distances.
Bounds  of the order of $M
> 10^{15} gev$ \cite{ellis} have been obtained from the non-observation of
an energy dependent time delay of Gamma Ray Burst signals.

An example of
a coupling term which does not give rise to an energy dependent
group velocity
is given by $L_I= \lambda~K^{\mu \nu \rho \sigma} ~F_{\mu \nu}
F_{\rho \sigma} $ \cite{kostalecky}, where $K^{\mu \nu \rho
\sigma}$ is a non-dynamical constant which has the same symmetries
under exchange of indices as the Riemann tensor. This coupling
violates $CPT$ and Lorentz invariance and it gives rise to
optical rotation of photon signals. Kostelecky and Mewes
\cite{kostalecky} put upper bounds $\sim 3 \times 10^{-32}$ on the
components of $K$ from the non-observation of optical activity in
signals from radio galaxies.

In the present example we find that due to the Riemann term, the
coupling becomes vanishingly small for cosmological background
metrics and the advantage of large cosmological distances over
which the effect of dispersion or polarisation is accumulated in
the examples discussed earlier are lost. We find that the
strongest bounds come from the consideration of compact stars.
Radar echo experiments of radio signals passing in the vicinity of
the sun give a bound of $\lambda < 3.9 \times 10^{19} cm^2$. This
is two orders of magnitude more stringent than bounds on $\lambda$
obtain from the birefringent bending of light by the sun
\cite{bedran}. We find that the most stringent bound on $\lambda$
is obtained from the timing of binary pulsar PSR B1534+12 signals
\cite{pulsar} and is given by $\lambda < 0.6\times 10^{11} cm^2$.

\section{Curvature coupling electrodynamics}

The equation of motion for $F^{\mu \nu}$ from the Lagrangian
(\ref{L}) is given by
\begin{equation}
\nabla_\nu F^{\mu \nu} = 2 \lambda \left[  R^{\mu \nu \rho \sigma}
 (\nabla_\nu F_{\rho \sigma}) + ( \nabla_\nu R^{\mu
\nu \rho \sigma}) F_{\rho \sigma} \right] \label{eom}
\end{equation}
 In
addition we have the Bianchi identities \begin{equation}
\nabla_\beta {R^{\mu \nu}}_{\rho \sigma}+ \nabla_\sigma {R^{\mu
\nu}}_{\beta \rho }+\nabla_\rho {R^{\mu \nu}}_{ \sigma \beta}=0
\label{bianchi} \end{equation}
  Setting $\beta = \nu$ in (\ref{bianchi}) we get
  \begin{equation}
  \nabla_\nu  {R^{\mu \nu}}_{\rho \sigma}=\nabla_\sigma
  {R^\mu}_\rho- \nabla_\rho {R^\mu} _\sigma
  \label{bianchi2}
  \end{equation}
 Using (\ref{bianchi2}) in the equation of motion (\ref{eom}) we
 obtain
 \begin{equation}
\nabla_\nu F^{\mu \nu} = 2 \lambda \left(  R^{\mu \nu \rho \sigma}
 (\nabla_\nu F_{\rho \sigma}) + ( \nabla_\rho R^\mu_\sigma)
  F^{\rho \sigma} \right) \label{eom2} \end{equation}
In the absence of sources ($ R^{\mu \nu}=0 $)   equation
(\ref{eom2}) simplifies to
\begin{equation}
 \nabla_\nu F^{\mu \nu} = 2 \lambda ~
R^{\mu \nu \rho \sigma}
 (\nabla_\nu F_{\rho \sigma})
 \label{eom3}
 \end{equation}

 To obtain the trajectory of photons in curved space it is convenient to choose a
 locally flat inertial frame and derive the dispersion relations
 for the photon momenta $p_{(\mu)}$ in the inertial frame. The experimentally observed
 photon momenta in the coordinate frame $k_\mu $ is then obtained
 by making use of the tetrads $k_\mu = e^{(\nu)}_\mu p_{(\nu)}$. In our
 application the photon wavelengths will be much smaller than the
 curvature scale of the gravitating bodies and therefore we can make
 the eikonal approximation
 \begin{equation}
 \nabla_{(\mu)} F^{(\mu)( \nu)} = p_{(\mu)} ~F^{(\mu)(\nu) }
 \end{equation}
 in which case the equations of motion (\ref{eom3}) in a local
 inertial frame appear as
 \begin{equation}
 p_{(\nu)} ~F^{(\mu)( \nu)} = 2 \lambda R^{(\mu)( \nu)( \alpha ) (\beta)}
p_{(\nu)}
 F_{(\alpha)(\beta)}
 \label{eom4}
 \end{equation}
 We also have the Bianchi identity,
 \begin{equation}
 p_{(\alpha)} F_{(\mu)(\nu)} + p_{(\nu)} F_{(\alpha)(\mu)} + p_{(\mu)}
F_{(\nu)( \alpha)} =0
  \label{bianchi4}
 \end{equation}
  Setting $\mu=j$ in (\ref{eom4}) ( we denote the time index by $0$
  and spatial indices by Latin  and four indices by Greek letters)
  we obtain
  \begin{eqnarray}
  p_{(0)} F^{(j) (0)} + p_{(k)} F^{(j)( k)} -
  2 \lambda (R^{(j) (\nu) (\alpha) (0)} p_{(\nu)} F_{(\alpha)(
  0)}
  + R^{(j) (\nu) (\alpha)( k) } p_{(\nu)} F_{(\alpha)( k)}) =0
  \label{eom5}
  \end{eqnarray}
  We use the Bianchi identity (\ref{bianchi4}) to write the magnetic
  field tensor in terms of electric field tensor
  \begin{equation}
  p_{(0)} F_{(j)(k)}= p_{(k)} F_{(j)(0)} - p_{(j)} F_{(k)(0)}
  \label{B}
  \end{equation}
 Using (\ref{B}) in (\ref{eom5}) we finally obtain the wave
 equation for the electric field polarization vector in a local
 inertial frame,
\begin{equation}
(p^{(\mu)} p_{(\mu)} \delta^{(j)}_{(k)} - p^{(j)} p_{(k)} +
4\lambda {R^{(j) (\nu) (\mu) }}_{(k)} p_{(\nu)} p_{(\mu)} ) ~F^{(k)( 0)}
=0
\label{eom6}
\end{equation}
Using (\ref{eom4}) to substitute for the second term we find that
(\ref{eom6}) reduces to the form
\begin{equation}
(p^{(\mu)} p_{(\mu)} \delta^{(j)}_{(k)} + 4 \lambda
(-\frac{p^{(j)}}{p_{(0)}} {\epsilon^{(0)}}_{(k)}+
{\epsilon^{(j)}}_{(k)}) F^{(k)( 0)} =0 \label{eom7}
 \end{equation}
where \begin{equation}
 {\epsilon^{(\alpha)}}_{(\beta )}\equiv {R^{(\alpha)( \mu)(
\nu)}}_{(\beta)} p_{(\mu)} p_{(\nu)}
\label{eps1}
  \end{equation}
  The
dispersion relations for electromagnetic waves can be obtained by
setting
\begin{equation} Det[p^{(\mu)} p_{(\mu)} \delta^{(j)}_{(k)} + 4 \lambda
(-\frac{p^{(j)}}{p_{(0)}} {\epsilon^{(0)}}_{(k)}+
{\epsilon^{(j)}}_{(k)})]=0 \label{det1}
\end{equation}

 For the Schwarzschild metric
 \begin{eqnarray} ds^2=
(1-\frac{ 2 G M}{r}) dt^2 -(1- \frac{2 G M}{r})^{-1} dr^2 -r^2 ( d
\theta^2 + Sin^2 \theta d \phi^2),
\end{eqnarray}
the non-zero components of Riemann tensor in a local inertial
frame are
\begin{eqnarray}
R_{(0)(1)(0)(1)}&=&-R_{(2)(3)(2)(3)}= \frac{2 G M}{r^3}\nonumber\\
R_{(0)(2)(0)(2)}&=& R_{(0)(3)(0)(3)}=
-R_{(1)(2)(1)(2)}=-R_{(1)(3)(1)(3)}=-\frac{G M}{r^3}
\end{eqnarray}
and the components of $\epsilon_{(\alpha)(\beta)}$ are
\begin{eqnarray}
\epsilon_{(0)(0)}&=& \frac{G M}{r^3}(-2 {p_{(1)}}^2+ {p_{(2)}}^2
+{p_{(3)}}^2 )\nonumber\\ \epsilon_{(1)(1)}&=& -\frac{G M}{r^3}(2
{p_{(0)}}^2+ {p_{(2)}}^2 +{p_{(3)}}^2 )\nonumber\\
\epsilon_{(2)(2)}&=& \frac{G M}{r^3}( {p_{(0)}}^2- {p_{(1)}}^2 + 2
{p_{(3)}}^2)\nonumber\\ \epsilon_{(3)(3)}&=& \frac{G M}{r^3}(
{p_{(0)}}^2- {p_{(1)}}^2 + 2 {p_{(2)}}^2 )\nonumber\\
\epsilon_{(0)(1)}&=&-\frac{2 G M}{r^3} p_{(0)} p_{(1)}\nonumber\\
\epsilon_{(0)(2)}&=&\frac{G M}{r^3} p_{(0)} p_{(2)}\nonumber\\
\epsilon_{(0)(3)}&=&\frac{G M}{r^3} p_{(0)} p_{(3)}\nonumber\\
\epsilon_{(1)(2)}&=&\frac{G M}{r^3} p_{(1)} p_{(2)} \nonumber\\
\epsilon_{(1)(3)}&=&\frac{G M}{r^3} p_{(1)} p_{(3)}\nonumber\\
 \epsilon_{(2)(3)}&=&-\frac{2 G M}{r^3}
p_{(2)} p_{(3)} \label{eps}
\end{eqnarray}
where we have set up the tetrad frame such that $(p_1,p_2,p_3)$
 are along $(\hat{e_r},\hat{e_\theta},\hat{e_\phi})$.

With these expressions we can determine the dispersion relation
(\ref{det1}) for the Schwarzschiled geometry which can be compared
with experimental observations.

\section{Polarization dependent time delay}

A radar signal sent across the solar system past the Sun to a
planet or a spacecraft suffers an additional Shapiro time delay in
general relativity which has been measured \cite{radar} and
confirms Einsteins theory to about one percent accuracy. In
presence of the curvature coupling terms there is an additional
time delay which we calculate in the following.
 We consider  photon trajectories tangential to
 the gravitating body and in the  tetrad frame
 $p_{(\mu)}=(p_0,p_1,0,p_3)$. The equation of motion for the $(E_1, E_2)$
components is of the form
\begin{equation}
\pmatrix{p^2 + c_1 ( 2(p_{(0)}^2-p_{(1)}^2) + p_{(3)}^2) & 0
\cr 0 & p^2-  c_1 ( p_{(0)}^2-p_{(1)}^2 +2  p_{(3)}^2)}
\pmatrix{E_1 \cr E_2}=0
\end{equation}
where $c_1=4 \lambda (GM/r^3)$. Setting the determinant equal to zero
yields the dispersion relations for the two propagating modes, 
\begin{equation}
p^2 = \pm 12 \lambda \frac{GM}{r^3} p_{3}^2
\end{equation}
 We obtain the dispersion relation in terms of the coordinate frame 4-momentum $k_\mu$
 by making the transformation $p_{(a)}={e^{\nu}}_{(a)}{k_{\nu}}$. The
 dispersion relations in terms of $k_\mu$ is
 \begin{eqnarray}
 (1-2GM/r)^{-1} k_t^2 &-&(1-2GM/r) k_r^2 -(1/r^2)
 k_\phi^2\nonumber\\
&\mp &12 \lambda \frac{GM}{r^3}\frac{k_\phi^2}{r^2})=0
 \end{eqnarray}
  At the point of inflection of the photon trajectory $k_r=0$ and
 one can evaluate the constant angular momentum $p_\phi$ in terms of the
 $k_0$ and the impact parameter $r_0$ as
\begin{equation}
k_\phi= r_0 k_t (1-2GM/r)^{-1/2}(1 \mp 12 \lambda \frac{
GM}{r_0^3})^{1/2} \label{kphi}
\end{equation}
Substituting (\ref{kphi}) in the  dispersion relation we get
\begin{eqnarray}
 (1-2GM/r)^{-1} k_t^2 &-&(1-2GM/r) k_r^2 -
k_t^2 \frac{r_0^2} {r^2}(1-2 GM/r)^{-1}\nonumber\\ &\mp&12 \lambda
k_t^2 \frac{GM}{r^2 r_0}(1-2 GM/r)^{-1} =0
 \end{eqnarray}

 We can now solve the dispersion relations for $k_t$ in terms
of $k_r$ to obtain,
\begin{equation}
k_t=\frac{k_r
(1-2GM/r)}{(1-\frac{r_0^2}{r^2}\frac{((1-2GM/r)}{(1-2GM/r_0)}(1-\lambda_\pm
(\frac{GM}{r0^3}))^{1/2}}
 \label{k0}
\end{equation}
where we have defined $\lambda_\pm= \pm 12 \lambda $. The
four-momenta are related to the coordinates as $k^t=dt/ds$ and
$k^r=dr/ds$ which enables us to obtain the photon trajectory
\begin{eqnarray}
\frac{dt}{dr}= \frac{g^{tt} k_t}{g^{rr} k_r } \label{k-x}
\end{eqnarray}
Using the relations (\ref{k-x}) and (\ref{k0}) we obtain the
relation between coordinate time $t$ and radius $r$ given by
\begin{eqnarray}
t_f-t_i=\int_{r_i}^{r_f} \frac{dr
(1-2GM/r)^{(-1)}}{(1-\frac{r_0^2}{r^2}\frac{((1-2GM/r)}{(1-2GM/r_0)}(1-\lambda_\pm
(\frac{GM}{r0^3}))^{1/2}} \label{tf}
\end{eqnarray}
Evaluating this integral, we obtain for the
time delay
\begin{eqnarray}
\Delta t_{if}&=&r_{if} + 2 GM Log(\frac{r_f^2 +\sqrt{r_f^2
-r_0^2}}{r_i^2 +\sqrt{r_i^2 -r_0^2}})\nonumber\\ &+&\lambda_\pm
\frac{GM}{r_0^2}(\frac{(r_f^2-r_0^2)^{1/2}}{r_f}
+\frac{(r_i^2-r_0^2)^{1/2}}{r_i})
\end{eqnarray}
The last term is polarization dependent and therefore if one sends
a pulse of unpolarized radio signal, the two orthogonal
polarizations have different arrival times at the earth and we
should observe a splitting of the pulse in time with the time lag
between the two polarizations given by \begin{eqnarray}
t_{+}-t_{-}&=&
(\lambda_+-\lambda_-)\frac{GM}{r_0^2}(\frac{(r_f^2-r_0^2)^{1/2}}{r_f}
+\frac{(r_i^2-r_0^2)^{1/2}}{r_i})\nonumber\\ &=&  \lambda~24~
\frac{GM_{sun}}{R_{sun}^2}(|Sin\theta_f| +
|Sin\theta_i|)\end{eqnarray} where $\theta_i$ is the angle between
the position of the earth and $\vec r_0$ and  $\theta_f$ is the
angle between the position of the planet/spacecraft and $\vec
r_0$. The timing measurements are accurate to $\sim 1 \mu sec$.
When the planet spacecraft is at superior conjunction ( $Sin
\theta_i=Sin\theta_f=1 $)and taking $M_{sun}=1.98\times 10^{33}
gm$ and $R_{sun}=6.95\times 10^{5} km$,  the lack of pulse
splitting of the observed radar signals to within a $\mu sec$
accuracy translates to a bound
\be
\lambda < 1.1 \times 10^{20} cm^2. \ee

This bound is about three orders of magnitude more stringent than
the one obtained in \cite{bedran} from the analysis of bending of
light by the gravitational field of the sun in non-minimally
coupled electro-magnetism.

 Since
the time is proportional to $GM/r^2$ this effect is largest for
compact stars where ($M\sim M_\odot$ and $R \sim 10 km$).
Observation of the time delay in binary pulsar PSR B1534+12
\cite{pulsar} yields a value of $6.3 \pm 1.3 \mu sec$ for the
Shapiro time delay by the binary companion which has a mass of $
1.33 M_\odot$ and radius of $10 km$. The theoretical estimate
\cite{theory} of this time delay assuming general relativity to be
the correct theory is $ 6.6 \mu sec$. Taking an upper limit of $1
\mu sec$ as the maximum contribution of the polarization dependent
correction to the GR time delay we have an get an upper bound on
$\lambda$ ,
\be
\lambda < 0.6 \times 10^{11} cm^2 \ee This bound is far more
stringent than those from the solar ranging experiments. Note that
the constraint on $\lambda$ from pulsars is obtained by taking the
difference of the travel times of the radio signals with
orthogonal polarizations. There are systematic uncertainties in
the measurement of total travel time from pulsars as discussed in
\cite{fairhead}. However these systematic effects like the
emission time, pulsar velocity, earths velocity and position etc
are independent of the polarization of the radio signal so these
uncertainties drop out when we take the difference of travel time
between two different polarizations.

For radial photons the four momentum in the inertial frame is
$p_{(\mu)}=(p_0,p_1,0,0)$ and only non-zero components of the
coordinate frame four momentum are $k_\mu= (k_t, k_r,0,0)$. The
determinant condition on radial propagation gives $p^2=0$ and the
non-minimal coupling terms have no effect on radially propagating
electromagnetic waves.

\section{Conclusions}

We find that the curvature coupling of electromagnetic fields
give rise to dispersion free photon propagation, where the photon phase
and group velocities are dependent on the polarization.
The best constraints on such couplings can be
obtained from the timing of electromagnetic waves in the vicinity of
compact objects. The curvature coupling situation is different from the
types of couplings
where the effect accumulates over the distance of propagation and
stringent bounds on extra equivalence principle violating couplings are
put from observations of radio galaxies \cite{kostalecky} or gamma-ray
bursts \cite{ellis}.

In this paper we have taken the view that the classical lagrangian
can possibly have EEP violating but general covariant terms
with large coefficients which we bound from observations.
If these curvature couplings arise as only as  loop corrections to
Einteins general relativity \cite{drumond} and if $\lambda \sim
{M_p}^{-2}$ then there are so far no realistic means of observing such
couplings in experiments.

\end{document}